\documentclass[prl,reprint,superscriptaddress]{revtex4-2}

\usepackage{bibentry}

 \usepackage{amssymb}
 \usepackage{mathrsfs}
 \usepackage{amsmath,amsfonts}
 \usepackage[normalem]{ulem} 
 \makeatletter
\newcommand*{\addFileDependency}[1]{
  \typeout{(#1)}
  \@addtofilelist{#1}
  \IfFileExists{#1}{}{\typeout{No file #1.}}
}
\makeatother

\usepackage{xr-hyper} 
\usepackage{hyperref}
\usepackage{graphicx}
\usepackage{float}

\usepackage[T1]{fontenc}
\usepackage{cancel}
\usepackage{soul}
\usepackage{color}
\usepackage{float}

\usepackage{mathtools}

\DeclarePairedDelimiterX\braket[2]{\langle}{\rangle}{#1 \delimsize\vert #2}
\usepackage{ulem}
\usepackage{titlesec}
 \titleformat{\paragraph}[hang]{\bfseries}{}{0pt}{\uline}
\titlespacing*{\paragraph}
{0pt}{3.25ex plus 1ex minus .2ex}{1.5ex plus .2ex}

\newcommand{\textcc}[1]{(#1)}

\usepackage{CJK}

\begin{document}
\begin{CJK*}{UTF8}{gbsn}
\title{ Spatio-Temporal Coupling Controlled Laser for Electron Acceleration}
\author{Lu Wang (汪璐)}
\email{lu.wangthz@outlook.com}
\affiliation{
Key Laboratory for Laser Plasmas, School of Physics and Astronomy, Shanghai Jiao Tong University,  China
}
\author{Uwe Niedermayer}
\affiliation{
Technische Universität Darmstadt, Schlossgartenstrasse 8, D-64289 Darmstadt, Germany
}

\author{Jingui Ma (马金贵)}
\affiliation{
Key Laboratory for Laser Plasmas, School of Physics and Astronomy, Shanghai Jiao Tong University,  China
}
\author{Weihao Liu （刘维浩）}
\affiliation{
College of Electronic and Information Engineering, Nanjing University of Aeronautics and Astronautics, China
}

\author{Dongfang Zhang （张东方）}
\email{dongfangzhang@sjtu.edu.cn}
\affiliation{
Key Laboratory for Laser Plasmas, School of Physics and Astronomy, Shanghai Jiao Tong University,  China
}

\author{Liejia Qian （钱列加）}
\affiliation{
Key Laboratory for Laser Plasmas, School of Physics and Astronomy, Shanghai Jiao Tong University,  China
}

\begin{abstract}
Limited by the difficulty in acceleration synchronization, it has been a long term challenge for on-chip dielectric laser-based accelerators (DLA) to bridge the gap between non-relativistic and relativistic regimes. Here, we propose a DLA based on a spatio-temporal coupling (STC) controlled laser pulse, which enables the acceleration of a non-relativistic electron to a sub-MeV level in a single acceleration structure (chirped spatial grating). It provides high precision temporal and spatial tuning of the driving laser via the dispersion manipulation, leading to a synchronous acceleration of the velocity increasing electrons over a large energy range. Additionally, the STC scheme is a general method and can be extended to driving fields of other wavelengths such as terahertz pulses. Our results bring new possibilities to MeV-scale portable electron sources and table-top acceleration experiments.   
\end{abstract}

\maketitle
\end{CJK*}
Particle accelerators have attracted a lot of interest over the past years ranging from medical imaging, therapy and fundamental sciences \cite{karzmark1984advances,atlas2012particle}. Radio frequency (RF)-powered devices are the conventional choice for the accelerating elements~\cite{lee2018accelerator}. However, its large size, high input power and costly infrastructures limit its utility and accessibility to broader scientific communities. The growing desires for on-chip accelerations,  portable medical devices, and radiotherapy machines motivate us to explore alternative technologies that are more compact and cost-effective \cite{sapra2020chip,england2014dielectric,peralta2013demonstration}. Recently multiple small scale novel accelerator concepts have been shown, such as laser-plasma accelerators, terahertz-driven accelerators and dielectric laser accelerators. Terahertz-driven accelerators show high degree of beam control, but are still limited by the available terahertz energy i.e. low field gradient (sub GV/m)~\cite{fulop2012generation,zhang2018segmented,curry2018meter,tibai2018relativistic}. Laser-plasma accelerators have shown extremely high field gradients on the order of 100 GV/m. However, it suffers from instability and difficulties in injection \cite{mangles2004monoenergetic,sprangle1988laser,wang2013quasi,faure2018review}.  Dielectric-laser accelerators (DLAs) \cite{england2014dielectric,mcneur2018elements} powered by femtosecond lasers is another promising option, owing to the high damage threshold in the dielectric material~\cite{stuart1995laser}, modern ultrashort pulse lasers, and nanofabrication technologies \cite{nezhadbadeh2020chirped}. It supports a few GV/m \cite{Soong2012laser} field gradient ($\sim$10 GV/m for Si$\text{O}_2$\cite{du1994laser} and $\sim$3 GV/m\cite{bonse2002femtosecond} for Si) inside a microstructure. 

\begin{figure}[t!]
\centering
\includegraphics[width=0.99\linewidth]{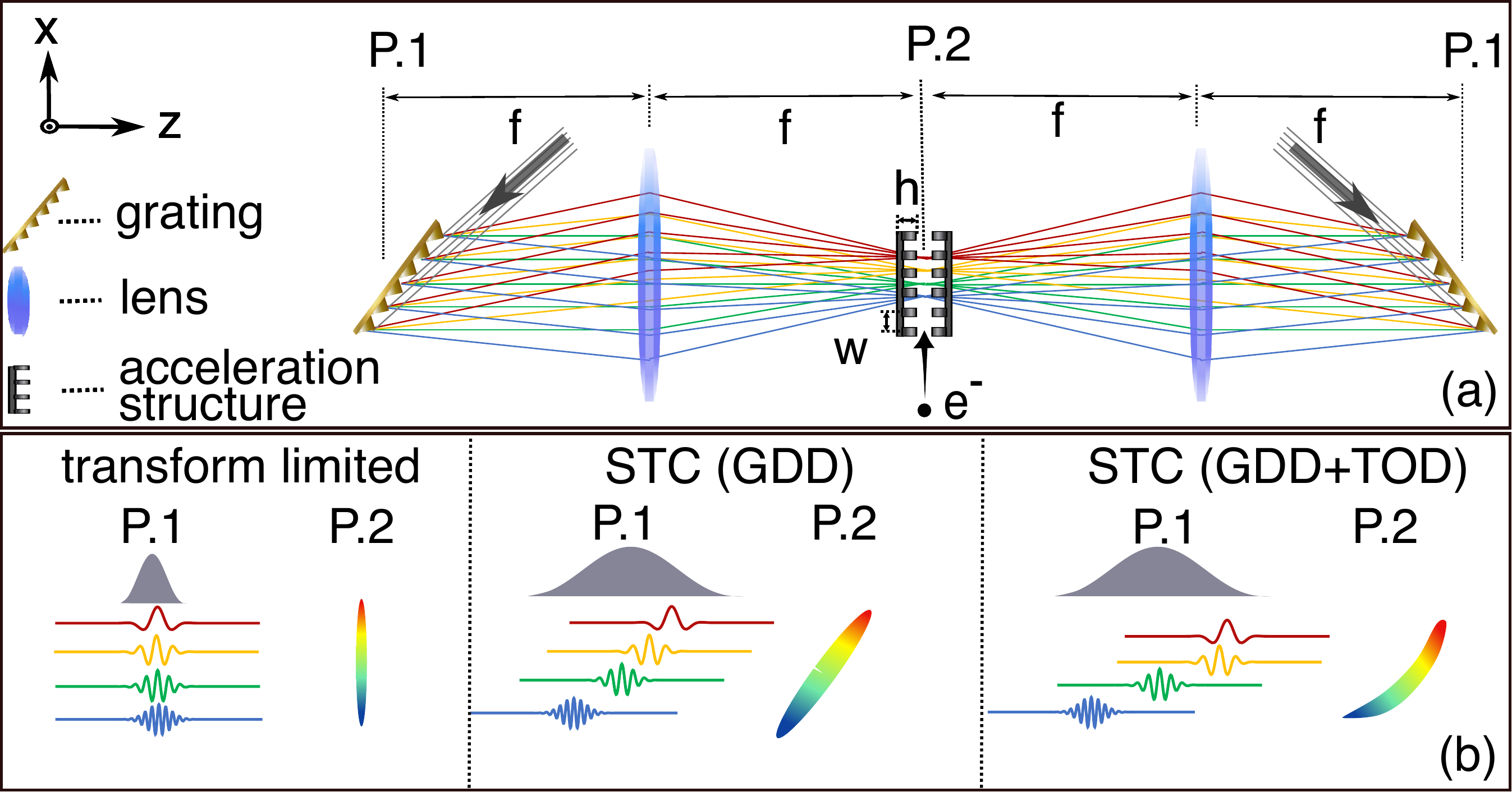}
\caption{Illustration of the spatio-temporal coupling (STC) scheme. \textcc{a} shows the illustration of the proposed acceleration configuration. In this symmetric configuration, two identical input laser pulses illuminate the grating simultaneously at position 1 (\textbf{P.1}). The electron interacts with counter-propagating fields at position 2 (\textbf{P.2}) in between the acceleration structures. The acceleration structure has period w and thickness h. \textcc{b} represents three different cases of input optical pulses. The temporal distribution at \textbf{P.1} converts to the spatial distribution along $x$ at \textbf{P.2}. }
\label{fig:illustration}
\end{figure}

Over the past 30 years, various setups have been proposed to optimize the acceleration process \cite{rosenzweig1995proposed,wei2017dual,kozak2017dielectric,cesar2018high,niedermayer2018alternating,niedermayer2021design}. Many fundamental functions required in an on-chip particle accelerator, such as acceleration, bunching, deflection and focusing have already been demonstrated experimentally using DLAs \cite{leedle2015laser,black2019laser,niedermayer2021low,shiloh2021electron}. However, it is still an remaining challenge to accelerate the electrons in the non-relativistic regime with ultrashort pulses. 
For uncorrelated, laterally impinging pulses, the electron acceleration length is restricted to the pulse duration $\tau$ (say $100$ fs), resulting in $v_0 \tau=6$ \textmu m with initial velocity $v_0=0.2c$. A pulse with a longer duration may be used to enlarge the interaction length, but this requires larger input energy at given electric field strength. The damage threshold fluence of the acceleration structure material prohibits such an approach or requires to resort to a lower field amplitude. 
A tree-network waveguide approach~\cite{hughes2018chip} can deliver short laser pulses at the right time and place, however, is overly complex for practical implementations. Currently among DLAs, the common approach to implement short laser pulses for a long acceleration length is by utilizing a pulse-front-tilted (PFT) laser pulse \cite{fulop2012generation,wootton2016demonstration,wei2017dual,kozak2017dielectric,cesar2018high,niedermayer2021design,niedermayer2020three}. The PFT scheme brings in a delay of the pulse along the particle acceleration direction $x$ (See Fig. \ref{fig:illustration} for coordinates definition),  making the short laser pulses at a given location $x$ arrive simultaneously with the electron. However, the PFT scheme can only match the driving laser with a fixed electron velocity, which fulfills $\tan\alpha=c/v$ and $\alpha$ represents the PFT angle. As a result, a walk-off occurs between the laser pulse and sub-relativistic electrons when the velocity increases due to acceleration.

To overcome the up-to-MeV DLA difficulty, a continuously changing, curved PFT is highly desired to obtain synchronous acceleration in conjunction with large speed increments. In this letter, we propose an all-optical-controlled spatio-temporal coupling (STC) controlled driving laser pulse, which changes its tilt angle according to the increasing velocity of the electron (see Fig. \ref{fig:illustration}). 
It is combined with a chirped dielectric structure~\cite{niedermayer2017designing}. The proposed scheme converts the temporal manipulation of the laser pulse into a spatially varying delay, which can be achieved by manipulating the group delay dispersion (GDD, $\Phi_2$) and third-order dispersion (TOD, $\Phi_3$). The STC scheme extends the interaction length and enhances the kinetic energy gain. Moreover, it retains a high flexibility in the optical operations for creating of the driving laser pulse.

The configuration we propose is shown in Fig. \ref{fig:illustration}\textcc{a}. Due to a symmetric setup with counter propagating (x-polarized) pulses, the magnetic fields cancel out at the channel center (\textbf{P.2}), and the electric fields add up. In Fig. \ref{fig:illustration}\textcc{b}, the sketches of three different cases of spectral phase induced PFTs are presented. 
\begin{figure}[b!]
\centering
\includegraphics[width=0.99\linewidth]{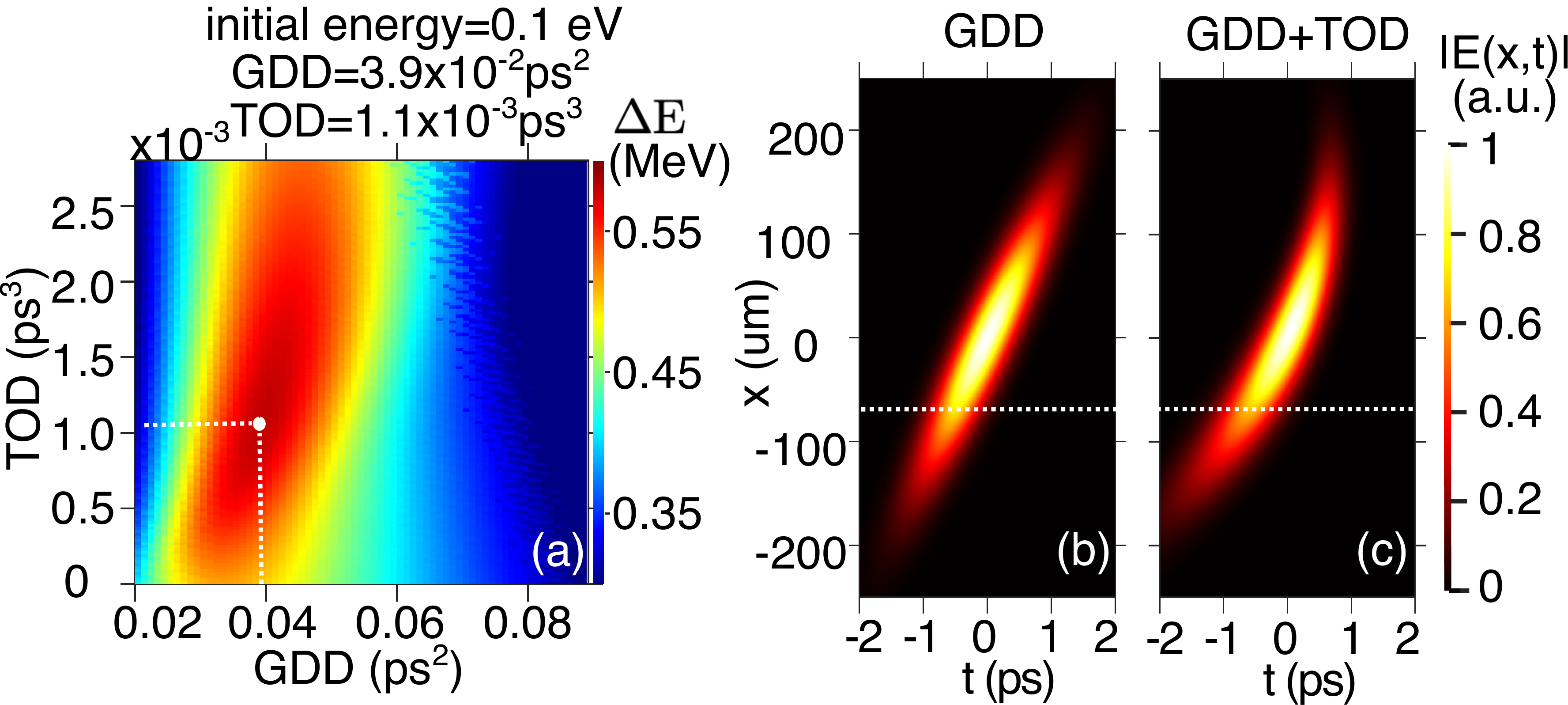}
\caption{Simulation for the perfectly matched pulse with 0.1 eV electron. (a) shows the kinetic energy gain as a function of the GDD ($\Phi_2$) and TOD ($\Phi_3$), where the maximum kinetic energy gain $\sim$0.6 MeV is represented by the white dot. With the parameters represented by he white dot, the peak field strength $E_0=2\times 2.4$ GV/m and a factor of 0.7 is considered for the evanescent field effect. The envelope of the electric fields $|E(x,t)|$ before interacting with the acceleration structure are presented in \textcc{b,c}, where $\Phi_2=3.9\times10^{-2}\text{ps}^2$, $\Phi_3=0$ and $1.1\times10^{-3}\text{ps}^3$, respectively. One can see that the GDD leads to a constant PFT along $x$ and TOD modifies the PFT along $x$. The white dashed line indicates the electron injection position $x=-0.45\sigma\prime_\text{FWHM}$, and $\sigma\prime_\text{FWHM}$ is the full-width-half-maximum of the beam size at location \textbf{P.2} in Fig.~\ref{fig:illustration}(a)}
\label{fig:matched}
\end{figure}

\begin{figure*}[htb]
\centering
\includegraphics[width=0.99\linewidth]{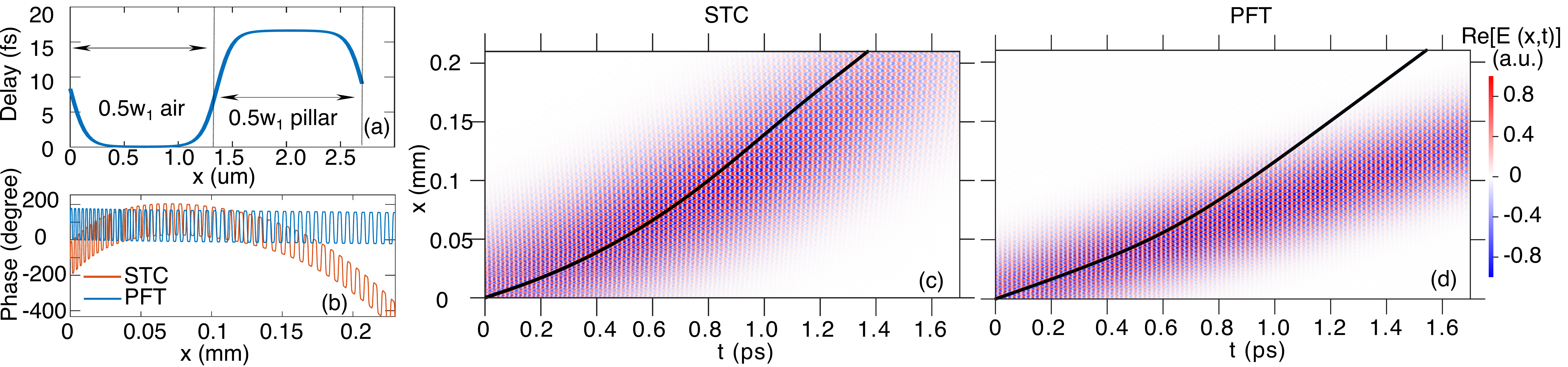}
\caption{Acceleration of a 20 keV electron with a specific acceleration structure. ({a}) shows the delay of the laser pulse induced by the first period of the acceleration structure. (b) shows the phase of the electric field along the electron trajectory as shown in (c,d), where the phase is defined as $\tan^{-1}\{$Im [$E(x,t)$]/Re[$E(x,t)$]$\}-\omega_0t$. ({c}) and (d) present the electric field distribution after the acceleration structure as a function of $x$ and $t$ for the STC and PFT schemes respectively. The black curves represent the electron trajectories. Note that the electric field presented is for location \textbf{P.2} with a fixed $z$. It can be seen that for the STC scheme, the electric field carries a curved phase-front, whereas the PFT scheme has a flat phase-front. }
\label{fig:E_compare}
\end{figure*}

\begin{figure*}[htb]
\centering
\includegraphics[width=0.9\linewidth]{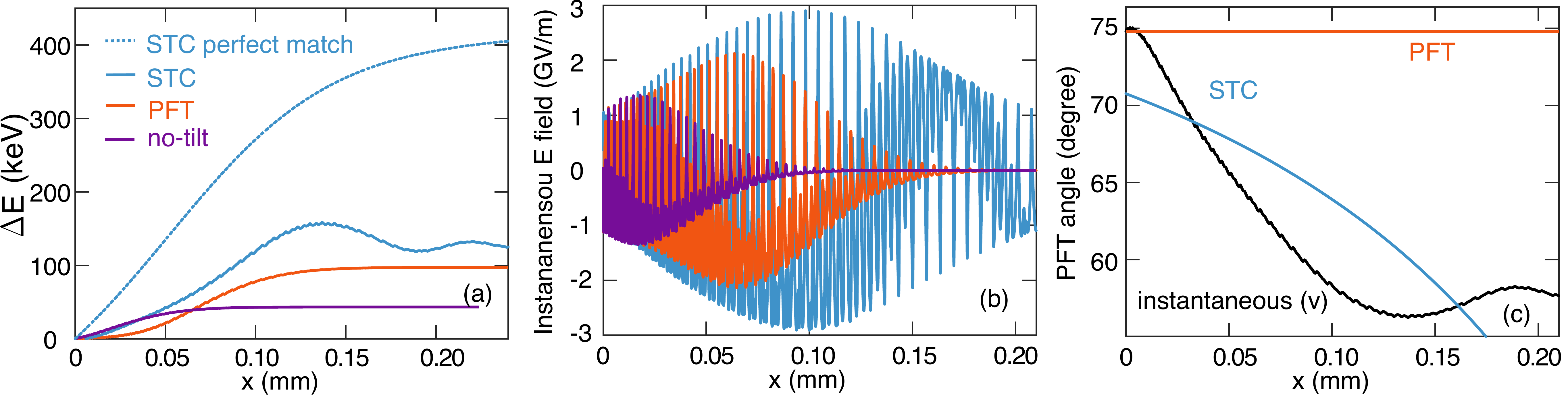}
\caption{Comparisons among the STC, PFT and no-tilt schemes for the electron with 20 keV initial kinetic energy. (a) shows the kinetic energy gain $\Delta E$ with peak field strength $E_0=2\times2.25~\text{GV}/\text{m}$. The maximum electric field for acceleration is $E_0$ multiplied with the evanescent field factor. The perfect-matched STC scheme with the same parameters is shown by the blue dashed curve as a reference. (b) shows instantaneous electric field applied on the electron (the field electron experiences during the acceleration). (c) shows the comparisons among the instantaneous PFT along the acceleration direction $x$, where the PFT scheme has a constant PFT while the STC scheme has a changing PFT. The instantaneous PFT angle is derived from the electron velocity. The parameters used are $\Phi_2=4.5\times10^{-2}\text{ps}^2,~\text{and}~\Phi_3=9.4\times10^{-4}\text{ps}^3$. At \textbf{P.2} the pulse duration $\tau\prime_\text{FWHM}=500$ fs, and the beam size $\sigma\prime_\text{FWHM}=0.16$ mm.}
\label{fig:e_field_PFTSTC}
\end{figure*}

In order to analyse the STC in detail. We chose to look into two aspects. One is the perfect matched situation shown in Fig.~\ref{fig:matched} where the electron with 0.1 eV initial kinetic energy, typical excess energy for photoelectron \cite{hauri2010intrinsic}, is assumed to sit on the envelope of the field. This corresponds to the ideal design of the acceleration structure where no dephasing between the driving field and the electron occurs. This gives us insights on the maximum kinetic energy gain with the given parameters. The other is the acceleration results of a 20 keV electron with a specific acceleration structure shown in Figs.~\ref{fig:E_compare} and ~\ref{fig:e_field_PFTSTC}. This gives a realistic guidance to future experimental work.

In Fig.~\ref{fig:matched}(a), the kinetic energy gain of a slow electron (0.1 eV) with a perfectly matched electric field is presented as a function of the GDD and TOD. A factor of 0.7 is included to take into consideration of the evanescent field effect. The maximum kinetic energy gain is $\sim 0.6$ MeV. It can be seen that the STC scheme is particularly advantageous for extremely low initial electron energies, i.e. large energy range. The electric field envelopes $|E(x,t)|$ are presented in \ref{fig:illustration}({c,d}), where the white dashed line represents the electron injection position $x=-0.45\sigma\prime_\text{FWHM}$ and $\sigma\prime_\text{FWHM}$ is the full-width-half-maximum of the beam size at \textbf{P.2}. Additionally, the electric fields at larger $t$ values ($t>0$) arrives later than that with smaller $t$ values ($t<0$). Note that for the convenience of the representation, in Fig.\ref{fig:illustration}({c,d}) $x=0$, $t=0$ is chosen to be where the peak intensity of the laser beam locates. In Figs.\ref{fig:E_compare} and \ref{fig:e_field_PFTSTC}, the initial electron acceleration location with the white dashed line is denoted as $x=0$. It can be seen that the GDD and TOD drastically influence the PFT shape. The TOD modifies the PFT along $x$, leading to a continuously matching PFT for the entire electron acceleration.  

For the STC scheme, the beam size and pulse duration at \textbf{P.2} largely depend on the parameters of the entire system.  We focus on a 2D+1 $(x,z,t)$ model where the electric fields in the frequency domain and the time domain are connected by Fourier transform $E(x,z,t)=\mathcal{F}[E(x,z,\omega)]$. Note that we use the complex notation for the electric fields and only the positive half of the spectrum is used i.e. $\omega>0$. The electric field used to calculate the electron acceleration is $\text{Re}[E(x,z,t)]$, where "Re" represents taking the real part. The incident electric field in the frequency domain before the grating at \textbf{P.1} follows the expression {\mbox{\cite{boyd2020nonlinear,keller}}}
\begin{align}\label{eq:e_in}
     E_\text{1}(x,0,\omega)=&A_{1}\exp{(-ikx^2/q_1)}\exp{(-\Delta \omega^2\tau^2/4)}\nonumber\\
   &\times \exp{[i(\Phi_2\Delta \omega ^2/2+\Phi_3\Delta \omega^3/6)]},
\end{align}
where $A_1$ is a constant representing the amplitude, $q_1=i\pi \sigma_1^2/\lambda_0$ is the q-parameter for a Gaussian pulse, $\sigma_1$ is the beam size, $k=2\pi/\lambda_0$ is the wave vector, $\lambda_0=10~$\textmu m is the center wavelength, $\tau_\text{FWHM}=100$ fs is the transform limited pulse duration (full-width-half-maximum), $\tau=\tau_\text{FWHM}/\sqrt{2\ln{2}}$, $\Delta \omega=\omega-\omega_0$, $\omega_0=2\pi c/\lambda_0$, $\Phi_2$ is the GDD, and $\Phi_3$ is the TOD. The electric field for electron acceleration at \textbf{P.2} is constructed by two steps. 

Firstly, the electric field reflects on the grating, propagates through the lens, and arrives at the acceleration structure. These are calculated analytically via the ABCDEF matrix method \cite{martinez1988matrix} (see SM Section IA 
). 
The analytical expression of the electric field right before the acceleration structure is shown as the following:
\begin{align}\label{eq:e_wx_out}
    E_\text{2}(x,2f,\omega)=&A_2\exp{[-ik(x-\beta\Delta \omega f)^2/q_2]}\exp{(-\tau^2\Delta \omega^2/4)}\nonumber\\
    &\times \exp{[i(\Phi_2\Delta \omega ^2/2+\Phi_3\Delta \omega^3/6)]},
\end{align}
where $\beta$ is the angular dispersion induced by grating at \textbf{P.1}, and $f$ is the focal length of the lens. The choice of $\beta$ can be found in SM Section IB.
The parameters $A_2$ and $q_2$ are amplitude and the q-parameter at \textbf{P.2}, which depend on $\beta$ and $f$ (see explicit expression in SM Section. IA
). Note that Eq. (\ref{eq:e_wx_out}), is the expression at the focal point i.e. the propagation distance after the lens is $f$. In principle, the propagation distance between the lens and the acceleration structure is $f-h$. We found that the extra propagation distance $h$ has a minor influence on acceleration results with our parameter choices since $f\sim$ cm and $h\sim \mu$m.
Moreover, the effect of different focal length due to the diffraction angle in Littrow configuration~\cite{kreier2012avoiding} can be neglected, since our pulse does not contain a large bandwidth. Thus, we present the electric field at \textbf{P.2} at the focal length as shown in Eq. (\ref{eq:e_wx_out}), due to its simple mathematical representation (see SM Section IA
for detailed analytical expression).

Secondly, each acceleration structure period $w_{n}$ is iteratively calculated with the electron acceleration process~\cite{niedermayer2017designing,niedermayer2017beam}. In other words, upon entering the acceleration structure, the electron velocity $v_0$ is used to calculate the first acceleration structure period, $\text{w}_\text{1}=\lambda_0 v_0/c$. The accelerator structure introduces a $x$-dependent delay onto the driving field as shown in Fig. \ref{fig:E_compare}({a}). Without the loss of generality, a smooth flat-top function is used as an approximation of the shape of the delay. With the acceleration structure material as silicon ($n_\text{Si}=3.5$) \cite{salzberg1957infrared} and initial kinetic energy 20 keV, the first period of the acceleration structure $\text{w}_1=2.7\mu$m. For each period of the acc, both vacuum and the pillar section take $50\%$ length of the entire period as shown in Fig.~\ref{fig:E_compare}({a}). 
In our work,  the maximum phase difference of the pulse at the vacuum and tooth/pillar regime within one period of the acceleration structure is taken as $\pi$, which is a very simplified design. The detailed optimization and design of the dielectric acceleration structure can be achieved by shifting the pillar positions and thickness. With the parameters of Figs.~\ref{fig:E_compare} and \ref{fig:e_field_PFTSTC}, acceleration results of the optimal design i.e. the perfect matched pulse are presented in Fig.S6~
in the SM. The specific design is beyond the focus of of this work and can be found in the work of Niedermayer et. al \mbox{\cite{niedermayer2018alternating,niedermayer2020three}}. The phase of the electric field that the electron experiences, i.e. phase deviation the structure needs to be designed to correct, is shown in Fig.~\ref{fig:E_compare}(b).

After traveling through distance $\text{w}_\text{1}$, the new electron velocity is used to calculate the next acceleration structure period $\text{w}_\text{2}$, and this process repeats till the end of the acceleration. The evanescent field effect of each period is calculated by $\exp{[-0.5l\sqrt{(2\pi/w_n)^2-k^2}]}$, where $l=1$ um is the gap distance between the two facing acceleration structures. The evanescent field decay factor varies from $\sim 0.3$ to $\sim 0.7$ in Fig. \ref{fig:E_compare}(c). In Fig.~\ref{fig:E_compare}({c,d}), the electric field along the acceleration direction $x$ versus the time $t$ at \textbf{P.2} (a fixed $z$) is plotted. It can be seen that the STC has a curved phase-front whereas the PFT scheme has a flat phase-front. Due to the continuously changing intensity front, the electron stays within the pulse in the STC scheme for the entire acceleration process. In contrast, for the PFT scheme, the electron walks off immediately with the pulse.

Figure~\ref{fig:e_field_PFTSTC} presents the comparisons among the STC scheme, PFT scheme, and the no-tilt (direct transverse injection without any pulse-front-tilt) scheme, where the acceleration results of the STC and PFT schemes are outcomes of the electric fields presented in Fig.~\ref{fig:E_compare}{c,d} respectively. For a fair comparison, the optical laser at \textbf{P.2} of the three schemes are chosen to have the same beam size and pulse duration. The kinetic energy gains are presented in  Fig.~\ref{fig:e_field_PFTSTC}(a). The perfect-matched STC scheme with the same parameters is shown by the blue dashed curve. The perfect matched case for all three cases can be found in Fig.~S6
~in SM. The peak electric field strength illuminating on the acceleration structure before considering the evanescent field effects is $2.25 ~\text{GV}/\text{m}$ from each side. Figure~\ref{fig:e_field_PFTSTC}(a) indicates that a matching PFT enhances the acceleration energy drastically. The STC scheme should show greater advantageous with higher acceleration field strength. The instantaneous electric fields the electron experiences along the acceleration position $x$ are shown in Fig. \ref{fig:e_field_PFTSTC}(b). It can be seen that for the PFT and no-tilt schemes, the electron walks off with the pulse imminently whereas, for the STC scheme, the electron sees the acceleration field for a longer interaction length. In Fig. \ref{fig:e_field_PFTSTC}({c}) the PFT angles are presented. The black curve is presented as a reference, where the instantaneous PFT angle is calculated from the electron velocity i.e. $\tan{(\text{angle})}=c/v(x)$. The PFT angle of the PFT scheme is a constant $c/v_0$.

In all the calculations presented in this letter, we assume a constant distribution along $y$ dimension with the beam size $\sigma_y=0.2$ mm and calculate the total input energy as $0.48\text{~mJ}=0.5c\varepsilon_0 \sigma_ y\iint |E(x,0,\omega)|^2dxd\omega=0.5c\varepsilon_0\sigma_ y\iint |E(x,2f,\omega)|^2dxd\omega$, where the $\varepsilon_0$ is the vacuum permittivity. 

The STC scheme enables high flexibility of the optical system elements. There is no constrain of the focal length, as long as the electron interaction point and the grating are positioned at each side of the lens' focal points. Additionally, this scheme enables high tunability since the PFT is defined by $\Phi_2$ and $\Phi_3$, which can be controlled by an commercially available acoustic-optical modulator. It offers independent programmable adjustment of GDD, TOD and higher-order dispersion on-the-fly. Meanwhile,  the adjustment of GDD and TOD does not influence the pre-aligned optical system. It provides possibility of fine adjustment of even higher order dispersion through electron feedback. Machine leaning \cite{lohani2019dispersion,genty2021machine} can also be implemented into the system to optimize the beam properties. Most importantly, the TOD modifies PFT along the $x$ dimension, resulting in a curved PFT that enables a continuously matching driving field to the electron beam for the entire acceleration process, which is crucial for high energy acceleration with short acceleration length. This largely enhances the flexibility of the experimental implementations.

To conclude, we present an all-optical-controlled scheme for non-relativistic electron acceleration in the DLA via the spatio-temporal coupling controlled driving pulse. It is promising especially for acceleration of non-relativistic electron with high electric field strength, where the electron velocity varies drastically during the acceleration process. The STC shows the possibility of high precision PFT angle control by converting the temporal variation into a spatial manipulation, which highly relaxes the nano-scale fabrication precision and increases the feasibility of implementing such a scheme with dielectric structures. 
Owing to the continuously matching PFT, STC provides long interaction length and high kinetic energy gain. The optical configuration enables unique continuous tunability of the optical intensity front shape by changing the GDD and TOD. The scheme is a general method which could potentially be applied to driving fields of other wavelengths. Our results bring new possibilities to portable electronic devices and table-top acceleration experiments.
\\


D.Z. gratefully acknowledge helpful discussions with Jing Wang, Liwen Zhang. L. W. thanks the HPC in Shanghai Jiao Tong University for allowing  $>$ 50 jobs parallel running. This brings great efficiency and happiness to the project. U.N. acknowledges funding by the Gordon and Betty Moore Foundation under Grant No. GBMF4744 (ACHIP), the German Federal Ministry of Education and Research (Grant No. 05K19RDE), and LOEWE Exploration. The work is supported by the National Natural Science Foundation of China (Grant No.12174255).\\
\\
The code developed for this work is publicly available upon request to the corresponding authors. 

\bibliography{main} 
\bibliographystyle{apsrev4-2}

\end{document}